\documentclass{aa} 
\usepackage{graphics}
\begin{document}

\title{The association between water kilomasers and compact radio sources in the starburst galaxy NGC~2146}

\author{A. Tarchi\inst{1,2}
        \and
        C. Henkel\inst{1}
        \and
        A. B. Peck\inst{1,3}
        \and
        K. M. Menten\inst{1}
        }

\offprints{A. Tarchi,
\email{atarchi@ira.bo.cnr.it}}

\institute{Max-Planck-Institut f\"ur Radioastronomie, Auf dem H\"ugel
              69, D-53121 Bonn, Germany
           \and
           Istituto di Radioastronomia, CNR, Via Gobetti 101, 40129 Bologna, 
              Italy
           \and
           Harvard--Smithsonian Center for Astrophysics, 60 Garden St., Cambridge MA 02138, USA 
}

\date{Received date / Accepted date}

\abstract{We report the detection of 22~GHz 
water vapor emission toward the starburst galaxy NGC~2146, made using the Effelsberg 100-m telescope. Interferometric 
observations with the Very Large Array (VLA) show that a part of the emission originates
from two prominent sites of star formation that are associated with compact radio 
continuum sources, likely ultra-compact HII regions. It is concluded that 
the emission arises from the most luminous and distant H$_2$O `kilomasers' detected so far. Our data increase the number of water maser detections in northern 
galaxies (Dec $>$--30$^{\circ}$) with IRAS point source fluxes $S_{\rm 100\mu m}$ 
$>$ 50\,Jy to 18\%.
}

\titlerunning{Water kilomaser in NGC~2146}
\authorrunning{Tarchi et al.}

\maketitle


\section{Introduction}

Interferometric studies of extragalactic water masers indicate that H$_{2}$O 
emission can arise in different environments. The weakest masers (with isotropic 
luminosities $L_{\rm H_{2}O} \rm < 0.1\,L_{\odot}$) are associated with prominent 
sites of star formation (e.g. in the LMC and SMC: Scalise \& Braz \cite{scalise82}; 
IC~342: Tarchi et al.\ \cite{tarchi02}), while the strongest ones (defined as 
`megamasers'; $L_{\rm H_{2}O} \rm > 20\,L_{\odot}$) are related to nuclear activity 
in the host galaxy, associated either with accretion disks (e.g.\ NGC~4258: Greenhill 
et al.\ \cite{greenhill95}) or with an interaction between the nuclear jet and 
molecular clouds (e.g.\ NGC~1052: Claussen et al.\ \cite{claussen98}; Mkn~348: 
Peck et al.\ \cite{peck01}). In the intermediate range of water maser luminosities 
($\rm 0.1\,L_{\odot} < \it {L_{\rm H_{2}O}} \rm < 20\,L_{\odot}$; the so called 
`kilomasers'\footnote{For a reference introducing this nomenclature, see 
e.g. Hagiwara et al.\ (\cite{hagi01}).}) there have been five extragalactic 
detections so far: M~33\,[IC~133] (Churchwell et al.\ \cite{church77}), 
M~82 (Claussen et al.\ \cite{claussen84}), IC~10 (Henkel et al.\ \cite{henkel86}),
and NGC~253 and M~51 (Ho et al.\ \cite{ho87}). Only the H$_{2}$O maser associated with 
M~51 (Hagiwara et al.\ \cite{hagi01}) has a nuclear origin, while most of the others 
are known to be associated with star-forming regions. Detecting new kilomasers and 
obtaining accurate positions is essential to search for the numerous population of nuclear 
kilomasers proposed by Ho et al.\ (\cite{ho87}), to identify regions of vigorous 
star formation in the outer disks, and to find targets that allow the determination
of proper motions.

In this paper we report on the detection of the most luminous and most distant 
H$_2$O kilomasers to date, and determine their positions with subarcsec 
accuracy toward the starburst galaxy NGC~2146.

\section{The observation and image processing}

{\bf Effelsberg} We observed NGC~2146 in the $6_{16} - 5_{23}$ transition of 
H$_2$O (rest frequency: 22.23508~GHz) with the 100-m telescope of the MPIfR at 
Effelsberg\footnote{The 100-m telescope at Effelsberg is operated by the 
Max-Planck-Institut f\"ur Radioastronomie (MPIfR) on behalf of the Max-Planck-Gesellschaft 
(MPG).} on April 6 and December 22, 2000. The beam width was 40$''$. The observations 
were made in a dual beam switching mode with a beam throw of 2\arcmin\ and a switching 
frequency of $\sim$1\,Hz. The system temperature, including atmospheric contributions, 
was $\sim$ 200 K (Apr.\ 2000) and $\sim$ 80 K (Dec.\ 2000) on a main beam temperature 
scale. Flux calibration was obtained by measuring W3(OH) (3.2\,Jy). Gain variations 
of the telescope as a function of elevation were taken into account. The pointing accuracy 
was always better than 10\arcsec.\\

\noindent{\bf VLA} NGC~2146 was observed on June 24, 2001 with the Very Large 
Array\footnote{The National Radio Astronomy Observatory is a facility of the National 
Science Foundation operated under cooperative agreement by Associated Universities, 
Inc.} (VLA) in its CnB configuration. We observed with two 25 MHz IFs centered on the 
two maser features shown in Fig.\,\ref{duemaser}. The source 0538+498 (1.93\,Jy) 
was used as flux calibrator. The point source 0718+793 was used for phase and bandpass 
calibration. The two IFs were calibrated separately and then `glued' together. The data 
were Fourier-transformed using natural weighting to create a $2048\times2048\times24$ data 
cube. The radio continuum was subtracted using the AIPS task UVLSF. This task fits a 
straight line to the real and imaginary parts of selected channels and subtracts the fitted 
baseline from the spectrum, optionally flagging data having excess residuals. In addition, 
it provides the fit continuum as a UV data set, which has been used to create the map shown 
in Fig.~\ref{2maservla}. The restoring beam is $0\farcs9\times0\farcs6$ and the rms noise 
per channel is 0.6 mJy, slightly higher than the expected thermal noise. This is likely 
caused by a cloudy sky during the observations\footnote{Sky coverage of 85\% with clouds 
of cumuloform type.}. 
 
\section{Results}

\begin{table*}
\caption{$\rm H_{2}O$ in NGC~2146: observational details and parameters of the two maser 
line components \label{fittare}}
\scriptsize
\begin{center}
\begin{tabular}{lccccccccc}
\hline
\\
Obs.\ Date & Tel.   & $\rm T_{int}^{a)}$ & BW    & Vel. Res.               & $\rm S_{peak}^{b)}$ 
& $\rm V_{LSR}^{c)}$ & $\Delta V_{\rm 1/2}$ & Luminosities$\rm ^{d)}$  \\
           &        & (hours)            & (MHz) & ($\mathrm{km\:s^{-1}}$) & (mJy)      
& ($\mathrm{km\:s^{-1}}$) & ($\mathrm{km\:s^{-1}}$) &  ${\rm \:L_{\odot}}$ \\
\hline
\\
Apr. 2000 &  EFF & 9 & 80 & 4.2 & 13   &  820 (3) & 70 (9)  & 4.3 \\
          &      &   &    &     & 13   & 1010 (3) & 60 (6)  & 4.0 \\
Dec. 2000 &  EFF & 2 & 80 & 4.2 & 10   &  825 (6) & 64 (19) & 3.0 \\
          &      &   &    &     & 10   & 1010 (6) & 94 (14) & 4.6 \\
Jun. 2001 &  VLA & 9 & 25 & 20  &  3.2 &  831 (2) & 33 (20) & 0.5\\
          &      &   &    &     &  6   & 1012 (2) & 41 (6)  & 1.2 \\
\\
\hline
\end{tabular}
\end{center}
a) $\rm T_{int}$ includes Effelsberg 22~GHz $\rm H_{2}O$ on- and  off-source integration 
time or VLA integration time including overheads\\
b) Peak fluxes calculated from integrated line intensities divided by linewidths (both 
obtained from Gaussian fits using the data reduction software package `CLASS')\\
c) LSR = Local Standard of Rest\\
d) Obtained via [$L_{\rm H_2O}$/L$_{\odot}$] = 0.023 $\times$ 
[$\int{S\,{\rm d}V}$/Jy\,km\,s$^{-1}$] $\times$ [$D$/Mpc]$^{2}$, $D$ = 14.5\,Mpc
\end{table*}

\begin{figure}[h]
\resizebox{8.0cm}{!}{\includegraphics{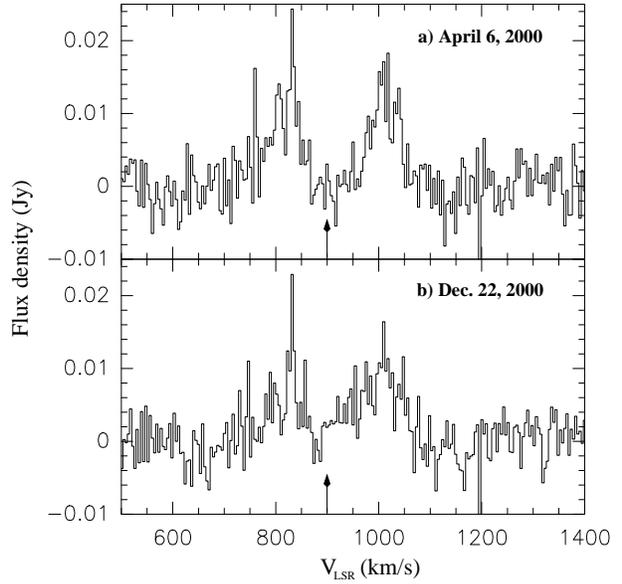}}
\caption{Water maser features in NGC~2146 observed with Effelsberg on {\bf a)} April 6, 2000, 
and {\bf b)} December 22, 2000. Arrows mark the systemic velocity ($\it{V}_{\rm LSR} =$ 
900 $\rm km\,s^{-1}$; De Vaucouleurs et al.\ \cite{devaucouleurs91}) \label{duemaser}}
\end{figure}

Fig.\,\ref{duemaser} shows two 22~GHz H$_2$O features, separated by $\sim$200\,km\,s$^{-1}$,
at velocities almost equidistant from the systemic velocity of NGC~2146 (see also 
Table~\ref{fittare}). The line width of each feature ($\sim$65$\rm \,km\,s^{-1}$) is 
larger than that of most other galactic or extragalactic maser components. The total 
isotropic luminosity of the emission is $\sim$8${\rm \,L_{\odot}}$ (see Table~\ref{fittare}; 
$D$=14.5 Mpc; Benvenuti et al.\ \cite{benvenuti75}). No other component was seen at velocities 
$\rm 380 < \it{V}_{\rm LSR} \rm < 1450 \:km\,s^{-1}$ (channel spacing: 4.2\,$\rm km\,s^{-1}$; 
$\rm 5\,\sigma$ noise level: 15 mJy).

\begin{figure}
\centering
\resizebox{8.5cm}{!}{\includegraphics{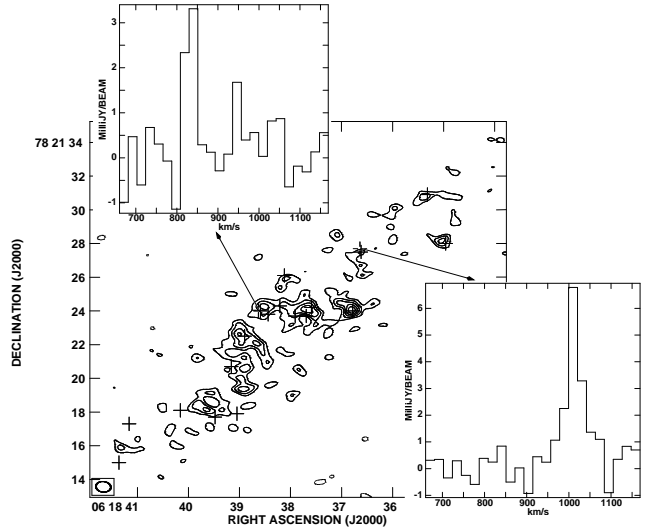}}
\caption{22~GHz continuum VLA contour plot of NGC~2146 (main panel) with the spectra 
of the two maser spots (small panels). The contour levels are $-$2.5, 2.5, 3.5, 4.5, 
5.5, 6.5, 7.5 $\sigma$ ($1\,\sigma$ = 0.2 mJy beam$^{-1}$). Crosses indicate the positions 
of the 18 compact radio sources (supernova remnants or HII regions) detected by TNG. 
The two crosses at the bases of the arrows mark the H$_2$O maser spots. The synthesized 
beam of 0\farcs9 $\times$ 0\farcs6 is shown in the lower-left corner. \label{2maservla}}
\end{figure}

Fig.\,\ref{2maservla} shows the 22~GHz VLA continuum map (central panel) obtained from 
the line-free channels as explained in Sect.~2 and the two spectra (small panels) 
of the maser emitting regions. Some compact radio sources are coincident with those detected 
at 5\,GHz by Tarchi et al.\ (\cite{tarchi00}; hereafter TNG), the positions of which are 
depicted by crosses. Differences between the map shown here and the 5~GHz map of TNG are 
attributable to the lower sensitivity and coarser positional accuracy of the former, and to 
the nonthermal nature of some sources (i.e. their radio flux density decreases with increasing 
frequencies).

\begin{figure}
\centering
\resizebox{8.5cm}{!}{\includegraphics{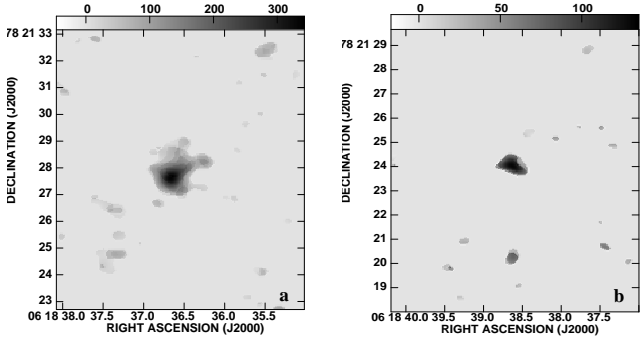}}
\caption{Total integrated intensity (in $\rm mJy\,beam^{-1}\,km\,s^{-1}$) VLA grey-scale plots 
of the ({\bf a}) high-velocity and ({\bf b}) low-velocity maser spots associated with the radio 
continuum sources 36.6+27.5 and 38.6+24.0, respectively. The synthesized beam is 0\farcs9 $\times$ 
0\farcs6.\label{2spots}}
\end{figure}

The water vapor emission seen by the VLA arises from two regions. The first one, 
at a position of \mbox{$\rm \alpha_{2000}$} = \mbox{06$\rm ^{h}$18$\rm ^{m}$36\fs64} 
and \mbox{$\rm \delta_{2000}$} = \mbox{78\degr21\arcmin27\farcs7}, is resolved and 
forms an extended structure with dimensions of $\sim$$\rm 2\arcsec \times 1.5\arcsec$ 
(140 $\times$ 105 pc; Fig.~\ref{2spots}a). The second region is located at \mbox{$\rm 
\alpha_{2000}$} = \mbox{06$\rm ^{h}$18$\rm ^{m}$38\fs63} and \mbox{$\rm 
\delta_{2000}$ = 78\degr21\arcmin24\farcs0} and is only slightly resolved with 
dimensions of $\sim$$\rm 1\arcsec \times 0.7\arcsec$ (70 $\times$ 50 pc; Fig.~\ref{2spots}b). Peak flux densities and restoring VLA beam size infer lower limits to the peak
brightness temperature of only $\sim$10\,K for the two H$_2$O lines. Since, however, the 22~GHz 
line is known to appear in all well studied sources as a maser, maser emission is also 
inferred in the following discussion for NGC~2146.

The maser positions coincide with two (36.6+27.5 and 38.6+24.0) of the eighteen continuum sources 
detected by TNG, which are also visible (in one case only barely) in the continuum image of 
Fig.~\ref{2maservla}. The positional errors, dominated by statistical errors, have been 
estimated using the synthesized beam size divided by the signal-to-noise ratios\footnote{For 
details on this derivation see e.g.\ Hagiwara et al.\ \cite{hagi01}}. The error on the 
continuum peak positions has been quadratically summed with that of the maser positions, 
giving a final value of 0\farcs16 for the positional uncertainty of both maser features. 
Within these uncertainties the continuum sources (36.6+27.5 and 38.6+24.0) and the maser 
spots are associated.

\section{Discussion}

The starburst galaxy NGC~2146 is characterized by large far infrared (FIR) and radio flux densities 
(e.g.\ Kronberg \& Biermann \cite{kronberg81}; Smith et al.\ \cite{smith95}) and by an outflow of 
hot gas along the minor axis driven by supernova explosions and stellar winds (Armus et al.\ 
\cite{armus95}, Della Ceca et al.\ \cite{dellaceca99}, Greve et al.\ \cite{greve00}). The star 
formation rate of NGC~2146, derived from its FIR flux density, is larger than that of the two 
starburst galaxies NGC\,253 and M\,82, which are also known to be H$_2$O masers sources (Ho et al.\ 
\cite{ho87}; Baudry \& Brouillet \cite{baudry96}).

\subsection{Nature of the water maser emission}

The nuclear position of NGC~2146 is near the compact radio source 37.6+24.2 (\mbox{$\rm \alpha_{2000}$ 
= 06$\rm ^{h}$18$\rm ^{m}$37\fs58}; \mbox{$\rm \delta_{2000}$ = 78\degr21\arcmin24\farcs2}; Tarchi et 
al.\ \cite{tarchi00}). The two maser spots are at a projected galactocentric distance of 7\arcsec\ 
(490 pc) and are not located symmetrically on either side of the nucleus. The maser spots are not part 
of the innermost region of the galaxy.

Based on the spectral index ($S_{\rm \nu} \propto \nu^{\alpha}$), the radio continuum source 
associated with the north-western maser (36.6+27.5) might be an optically thick compact HII region 
containing a number of massive stars (see TNG). A similar nature is also hypothesized for 
38.6+24.0 based on its detection at 5~GHz but not at 1.6~GHz, implying a steep positive spectral index 
$\alpha$. The velocity of the two features is consistent with the velocity field of the central CO 
concentration, described as a warped molecular ring with an extent of $\sim$1\,kpc (Greve et al. 
\cite{greve00}). The observed non-nuclear positions, the associated radio continuum sources, and the 
known connection between CO-emission and star formation indicate that the masers in NGC~2146 are related 
to sites of massive star formation and do not provide new evidence for the presence of a class of nuclear H$_2$O 22\,GHz kilomasers that was postulated by Ho et al. (\cite{ho87}). 

Integrated flux densities, peak intensities, and linewidths of the masers as derived from the VLA 
observations are smaller than the corresponding values from the single-dish detection (see 
Table~\ref{fittare}). This could be due to time variability of the masers. The fact that the 
measured single-dish profiles are broad and look similar in April and December 2000 suggest, 
however, that each of the two main velocity components represents a large number of individual 
subcomponents whose intrinsic variations almost cancel out. This is further supported by the spatial 
extensions of the maser spots (see Fig.\,\ref{2spots} and Sect.\,3) that suggest that the emission is 
the result of an overlap of several unresolved masers of which only the strongest and most compact 
sources (36.6+27.5 and 38.6+24.0) have been detected. While the VLA data allow us to locate the two 
most prominent groups of maser features with modest signal-to-noise ratios, many additional features 
may be too weak to detect. Within this context it is interesting that continuum emission near the 
two sources mentioned is seen both in the 22 and 5~GHz continuum maps (Fig.\,\ref{2maservla}, TNG).

\subsection{Where to search for water maser emission?}

The new detection of water maser emission in NGC~2146 increases to 18\,\% (8 out of 44) the detection 
rate of the complete sample of galaxies selected on the basis of their declination \mbox{($\rm \delta 
> -30^{\circ}$)} and IRAS (Infrared Astronomy Satellite) fluxes ($S_{\rm 100\mu m}$$>$50\,Jy; see 
Henkel et al.\ \cite{henkel86}; Tarchi et al.\ \cite{tarchi02}). If we consider the subsample 
of galaxies having $\rm S_{100\mu m}$$>$100\,Jy, which includes NGC~2146, the detection rate is 
$\sim$30\% (6 out of 19). Even taking into consideration that we are still dealing with small number 
statistics, the sample represents a unique case since typical searches for extragalactic 
water maser emission yield detection rates below (usually well below) 10\,\% (e.g.\ Henkel et al.\ 
\cite{henkel84}, \cite{henkel98}; Braatz et al.\ \cite{braatz96}; Greenhill et al.\ \cite{greenhill02}). 
The high detection rate may be more due to the proximity of the galaxies surveyed rather than due to 
a physical relationship between maser and IRAS far-infrared emission (Greenhill \cite{greenhill01}). 

Within a radius of 20\,Mpc, 16 external galaxies are known to emit 22\,GHz H$_2$O emission 
(marginal detections that are not included in our compilation were reported by Huchtmeier et al.\ 
\cite{huchtmeier78}, \cite{huchtmeier88}). There are seven megamasers with nuclear origin (NGC~1052, 
NGC~1068, NGC~1386, NGC~3079, NGC~4258, NGC~4945, Circinus; all classified as Sy\,2s and/or LINERs, 
and all spirals except the elliptical NGC~1052), 3 are low-luminosity `galactic-type' masers (the LMC, 
the SMC, and IC~342), and the remaining six sources (M~33\,[IC133], IC~10, NGC~253, NGC~2146, M~82, 
M~51) host kilomasers with average flux densities equal to or larger than the strongest galactic masers. 
Interferometric studies have shown that the masers in M~33\,[IC133], IC~10, and M~82 (all spirals or 
irregulars) are associated with off-nuclear star-forming regions. Hagiwara et al.\ (\cite{hagi01}) found 
that the H$_{2}$O emission in M~51 (spiral, HII, Sey2.5) is nuclear in origin, possibly associated with 
the receding jet. For NGC~253 (spiral, starburst) interferometric studies are not yet published.

So far, the less luminous masers associated with star-forming regions have all been detected at 
distances $<$4~Mpc, while all the luminous masers associated with nuclear activity, and hosted by AGNs, 
have been found at greater distances. The masers detected in NGC~2146 are the first with non-nuclear 
origin and $D$$>$4\,Mpc. These results point once more to the necessity of deep searches for 
H$_2$O maser emission in galaxies with large IRAS flux densities. These have been shown to possess a 
particularly large number of detectable masers from star forming regions that might be excellent targets 
to determine not only radial velocities but also galactic proper motions and distances.

\begin{acknowledgements}

We wish to thank M. Claussen, Y. Hagiwara and E. Ros for useful comments and discussions.
We are also endebted to the operators at the 100-m telescope, and to M. Rupen and the NRAO analysts, 
for their cheerful assistance with the observations. 
\end{acknowledgements}


\begin{thebibliography}{}
\bibitem[1995]{armus95}
 Armus, L., Heckman, T.M., Weaver, K.A., \& Lehnert, M.D.\ 1995, ApJ, 445, 666
\bibitem[1996]{baudry96}
 Baudry, A., \& Brouillet, N.\ 1996, A\&A, 316, 188
\bibitem[1975]{benvenuti75}
 Benvenuti, P., Capaccioli, M., \& D'Odorico, S.\ 1975, A\&A, 41, 91
\bibitem[1996]{braatz96}
 Braatz, J.A., Wilson, A.S., \& Henkel, C.\ 1996, ApJS, 106, 51
\bibitem[1977]{church77}
 Churchwell, E., Witzel, A., Pauliny-Toth, I., et al.\ 1977, A\&A 54, 969
\bibitem[1984]{claussen84}
 Claussen, M.J., Heiligman, G.M., \& Lo, K.Y.\ 1984, Nature, 310, 298  
\bibitem[1998]{claussen98}
 Claussen, M.J., Diamond, P.J., Braatz, J.A., Wilson, A.S., \& Henkel, C.\ 1998, ApJ, 500, L129
\bibitem[1999]{dellaceca99}
 Della Ceca, R., Griffiths, R.E., Heckman, T.M., Lehnert, M.D., \& Weaver, K.A.\ 1999, ApJ, 514, 772
\bibitem[1991]{devaucouleurs91}
 De Vaucouleurs, G., De Vaucouleurs, A., Corwin Jr., et al.\ 1991, Third Reference Catalogue of 
 Bright Galaxies, Version 3.9
\bibitem[2001]{greenhill01}
 Greenhill L.J., to appear in the proceedings of IAU Symposium 206: Cosmic Masers (Mangaratiba, 
 Brazil; March 2001) -- astro-ph/0109419
\bibitem[1995]{greenhill95}
 Greenhill, L.J., Jiang, D.R., Moran, J.M., et al.\ 1995, ApJ 440, 619
\bibitem[2002]{greenhill02}
 Greenhill, L.J., Ellingsen, S.P., Norris, R.P., et al.\ 2002, ApJ 565, 836
\bibitem[2000]{greve00}
 Greve, A., Neininger, N., Tarchi, A., \& Sievers, A.\ 2000, A\&A, 364, 409
\bibitem[2001]{hagi01}   
 Hagiwara, Y., Henkel, C., Menten, K., \& Nakai, N.\ 2001, ApJ, 560, L37
\bibitem[1984]{henkel84}
 Henkel, C., G{\"u}sten, R., Wilson, T.L., et al.\ 1984, A\&A, 141, L1
\bibitem[1986]{henkel86}
 Henkel, C., Wouterloot, J.G.A., \& Bally, J.\ 1986, A\&A, 155, 193
\bibitem[1998]{henkel98}
 Henkel, C., Wang, Y.P., Falcke, H., Wilson, A.S., \& Braatz, J.A.\ 1998, A\&A, 335, 463
\bibitem[1987]{ho87}
 Ho, P.T.P., Martin, R.N., Henkel, C., \& Turner, J.L. \ 1987, ApJ, 320, 663
\bibitem[1978]{huchtmeier78} 
 Huchtmeier, W.K., Witzel, A., K\"uhr, H., Pauliny-Toth, I.I, \& Roland, J.\ 1978, A\&A, 64, L\,21
\bibitem[1988]{huchtmeier88}
 Huchtmeier, W.K., Eckart, A., \& Zensus, A.J.\ 1988, A\&A, 200, 26
\bibitem[1981]{kronberg81}
 Kronberg, P.P., \& Biermann, P.\ 1981, ApJ, 243, 89
\bibitem[2001]{peck01}
 Peck, A.B., Falcke, H., Henkel, C., \& Menten, K.M.\ ASP Conf.\ Proc.\ 249, The Central Kiloparsec 
 of Starbursts and AGN, eds. J.H. Knapen et al., San Francisco, 2001, p.\ 321
\bibitem[1982]{scalise82}
 Scalise, E.Jr., \& Braz, M.A.\ 1982, AJ, 87, 528
\bibitem[1995]{smith95}
 Smith, B.J., Harvey, P.M., \& Lester, D.F.\ 1995, ApJ, 442, 610
\bibitem[2000]{tarchi00}
 Tarchi, A., Neininger, N., Greve, A., et al. 2000, A\&A, 358, 95 (TNG)
\bibitem[2002]{tarchi02}
 Tarchi, A., Henkel, C., Peck, A.B., \& Menten, K.M.\ 2002, A\&A, 385, 1049 
\end{thebibliography}
\end{document}